# Direct Digital-to-Physical Synthesis: From mmWave Transmitter to Qubit Control

Najme Ebrahimi, *Member, IEEE*, Haoling Li, *Graduate Student Member, IEEE*, Gun Suer , Kin Chung Fong , *IEEE Member*, and Leonardo Ranzani, *Senior Member, IEEE*

*Abstract-* The increasing demand for high-speed wireless connectivity and scalable quantum information processing has driven parallel advancements in millimeter-wave (MMW) communication transmitters and cryogenic qubit controllers. Despite serving different applications, both systems rely on the precise generation of radio frequency (RF) waveforms with stringent requirements on spectral purity, timing, and amplitude control. Recent architecture eliminates conventional methods by embedding digital signal generation and processing directly into the RF path, transforming digital bits into physical waveforms for either electromagnetic transmission or quantum state control. This article presents a unified analysis of direct-digital modulation techniques across both domains, showing the synergy and similarities between these two domains. The article also focuses on four core architectures: Cartesian I/Q, Polar, RF-Digital-to-Analog Converter (DAC), and harmonic/subharmonic modulation across both domains. We analyze their respective trade-offs in energy efficiency, signal integrity, waveform synthesis, error mitigations, and highlight how architectural innovations in one domain can accelerate progress in the other.

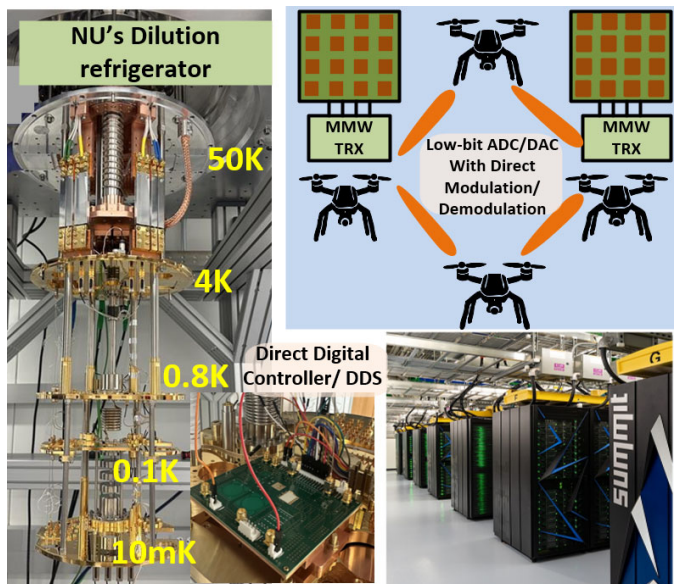

**Fig.1**: Conceptual view of next-generation digital-to-physical systems linking mm-wave communication, autonomous drones/UAVs, data-center interconnects, and cryogenic quantum processors for scalable waveform generation and control.

## I. INTRODUCTION

THE next-generation of MMW and sub terahertz (sub-THz), communication and computing nodes require versatile connectivity from data centers to edge-to-edge communications, Fig. 1. On the communications side, the applications for such high-speed connectivity are high-capacity links for data-center racks, edge-to-edge or drone-to-drone connections, and immersive device streams that demand both bandwidth and energy efficiency, Fig. 1. Both data-center communication, and future-generation quantum computation and sensing need high data rates (Tb/sec), enabled by advanced optical and electronic interconnects in the former and quantum devices in the latter [1-11]. For the shorter range and edge to edge connectivity, high-speed wireless node employs direct on-off keying (OOK), amplitude/phase shift keying (ASK/PSK/APSK) towards higher order quadrature amplitude modulation (QAM). The next generation of cryogenic qubit control circuits, [12-21] as well as MMW transmitters, [22-40], share structural similarities and mechanisms in how hardware architecture and non-idealities impact the signal integrity and errors. The new direct digital electronics moves the processing from bits to physical fields, [22-40], from the symbols produced by digital base band (BB) circuitry to the physical waveforms that either radiate through space via electromagnetic waveforms or rotate the state of a quantum device. This article focuses on the next generation of electronic drivers and circuitry that convert digital information directly into physical phenomena, enabling the transformation of bits into electromagnetic waves radiated from antenna elements for communication, or into precise microwave pulses that control qubit rotation angles and amplitudes in quantum systems. Specifically, it provides a comprehensive analysis of state-of-the-art (SoA) direct-modulation MMW transmitters and cryogenic quantum controllers for superconducting qubits. We anticipate a strong synergy in the development of the direct digital-to-quantum synthesis and MMW direct digital modulator for these two rising technological domains.

For MMW and sub-THz application, the goal is to bring the constellation and waveform generation/processing right at RF/MMW. Direct-digital schemes push the signal processing into fast switches or digitally weighted unit cells, so the transceivers no longer rely on multi-bit, multi-GS/s digital to analog converter (DACs) or analog-to-digital converter (ADC) and their high-speed clocks. Such solutions mitigate the size, power, and complexity of multi-bit, multi-GS/s DAC/ADC chains [22-40]. In recent SoAs, direct digital transmitters synthesize 16-QAM at the carrier, [33], [34], which can be extended to 64-QAM [34], using Cartesian or Polar based architecture. Alternative switch-based designs can generate 16-QAM, APSK, or $M$-ASK without a Nyquist-rate baseband [34], [24-26], [39]. These transmitters develop data-dependent distortions caused by fast on–off switching, I/Q amplitude and phase imbalance arising from state-dependent impedance changes, and local oscillator (LO) feedthrough or asymmetry that together increase the error vector magnitude (EVM). As data rates rise and constellations become denser, these imperfections intensify. Therefore, different architecture, (Cartesian, Polar, RFDAC, harmonic/subharmonic modulator) and circuit design innovations are required. Modern designs typically make use of multiple innovations at the circuit and architectural level with fine on-off leakage, I/Q and LO-feed through cancellation [22-39]. In addition, direct calibration and digital correction are performed close to each block with digital predistortion, pulse shaping, and light equalization.

A comparable development is evolving in quantum control systems. Despite advancements in superconducting qubit performance, operating under extreme cryogenic conditions (e.g., 10 mK), modern qubits still exhibit gate error rates around $10^{-4}$ or higher for two-qubit gates [1]. Building a useful fault-tolerant quantum computer which projected to consist of the order of one million physical qubits, demands a fundamental innovation of the control architecture. The challenge is transitioning the control and measurement infrastructure from racks of room-temperature hardware to compact integrated electronics. Moreover, wiring between the room temperature electronics and the qubit system leads to severe constraints on system size, because of the limited space and cooling power of refrigeration systems. One promising approach consists in positioning part of the electronics in the cryogenic environment, near the qubit array, together with other innovative multiplexing/demultiplexing approaches.

On the quantum side, the case for direct-digital modulation for qubit control mirrors its appeal from RF/MMW systems: precision, compactness, and integration. Gate fidelity is highly susceptible to amplitude, phase, and timing errors. As a result, the most robust path forward involves synthesizing exactly the desired waveform, including both carrier and envelope, directly on-chip using digital-to-analog converters [12-14], [2], direct digital synthesis (DDS) [18-19], [3], and direct RF modulators [16], [17]. These improvements translate to better long-term stability and gate fidelity. However, like MMW direct radios, pushing digital control deeper into the RF domain introduces nonlinearities such as on-off leakage, I/Q mismatches, LO feedthrough, and other data-dependent distortion, that can impact signal quality. Therefore, architectural innovation (Cartesian, Polar, RFDAC, harmonic/subharmonic controller), circuit innovations, calibration, and error analysis become essential for error mitigations and maintaining system fidelity at scale. *We therefore expect that advancement in one field will directly benefit the other field.*

In this article, we present a unified comparative framework that connects direct digital-to-physical synthesis challenges and trade-offs across MMW transmitters and cryogenic qubit-control architectures. Section II begins with a systematic analysis of domain-specific performance metrics, namely, EVM in MMW communication and gate fidelity in quantum computing and maps their underlying correspondence. Section III and IV introduce state-of-the-art circuit and architectural innovations, spanning I/Q (Cartesian) modulators, polar transmitters, direct RF-DACs, and harmonic/subharmonic-based mixer-modulators, for both RF/MMW signal transmission and modulation and qubit pulse control. We examine the key nonidealities and circuit-level impairments in each architecture, from direct QAM generation to vector-modulated qubit pulses as well as state-of-the-art performance. Finally, we conclude in Sec. V that the cross-domain analysis reveals a convergence in design objectives for next-generation electronics. Both communication and quantum systems require fine-grained amplitude and phase control, spectral purity, and robust error mitigation. Across domains, similar techniques, such as quadrature LO generation, calibration, LO leakage and spur suppression, impedance stabilization, pulse shaping, and harmonic cancellation, are employed to achieve stringent performance targets, including EVM < –20 dB for 5G/6G systems and gate error < $10^{-4}$ for fault-tolerant quantum computing.

## II: SYSTEM ANALYSIS FROM COMMUNICATION CONSTELLATION TO QUBIT CONTROL

Controlling a qubit or transmitting a direct MMW communication signal both involve generating precise RF waveforms, however the requirements are different. Qubit control pulses are short (ns–µs) microwave bursts designed to rotate a two-level system on the Bloch sphere with clean envelopes to avoid exciting unwanted spurious tones and unwanted qubit transitions. In contrast, MMW communication signals carry high-rate data (e.g. QAM constellations) and must meet spectral masks and error vector magnitude specifications and energy efficiency. Fig.2 (a) shows a qubit controller diagram with integrated Josephson Junctions, while Fig. 2 (b) presents an MMW chip with antenna array for direct MMW modulation and transmission. These systems can scale to support large qubit arrays or high-throughput multiple input multiple output (MIMO) transmitters, shown in Fig. 2. Each qubit or antenna array connects to a dedicated integrated circuit (IC) that provides I/Q or vector modulation, facilitating high-fidelity qubit control or direct RF waveform synthesis for communication constellations.

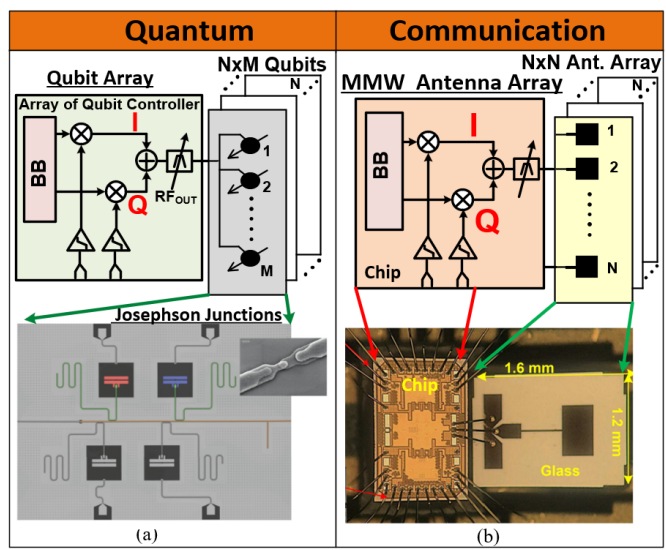

**Fig. 2:** Array-level analogy between qubit control and MMW transmitter. (a) Qubit controller driving an $N \times M$ qubit array, with an example qubit chip incorporating Josephson-Junction devices in [9]. (b) MMW direct-transmitter channel driving an $N \times N$ antenna array, with an example of chip integrated with an antenna-on-glass array [26]. In both systems, each channel synthesizes an I/Q (vector) RF waveform that is replicated across elements.

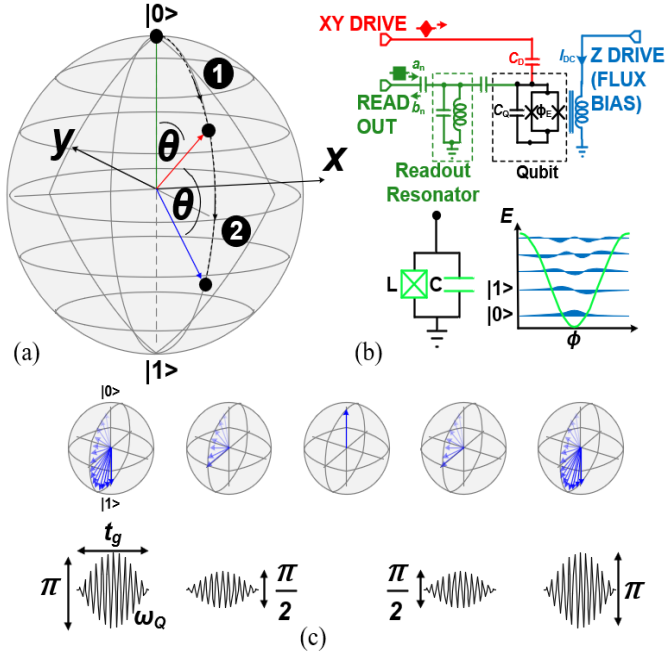

**Fig. 3:** Qubit control waveforms and physical interpretation. **(a)** Bloch-sphere view of single-qubit rotations: an on-resonance microwave pulse drives rotations between |0⟩ and |1⟩, where the pulse area sets the rotation angle $\theta$ and the carrier phase sets the rotation axis in the XY plane, [12] **(b)** Simplified superconducting-qubit control stack showing the XY drive (capacitive coupling), Z flux-bias path, and resonator-based readout used for state manipulation and measurement, [12] **(c)** Example gate pulses and resulting state evolution, illustrating $\pi$- and $\pi/2$-rotations over a gate time $t_g$ at carrier frequency $\omega_q$ of RF waveform that is replicated across many elements, [14].

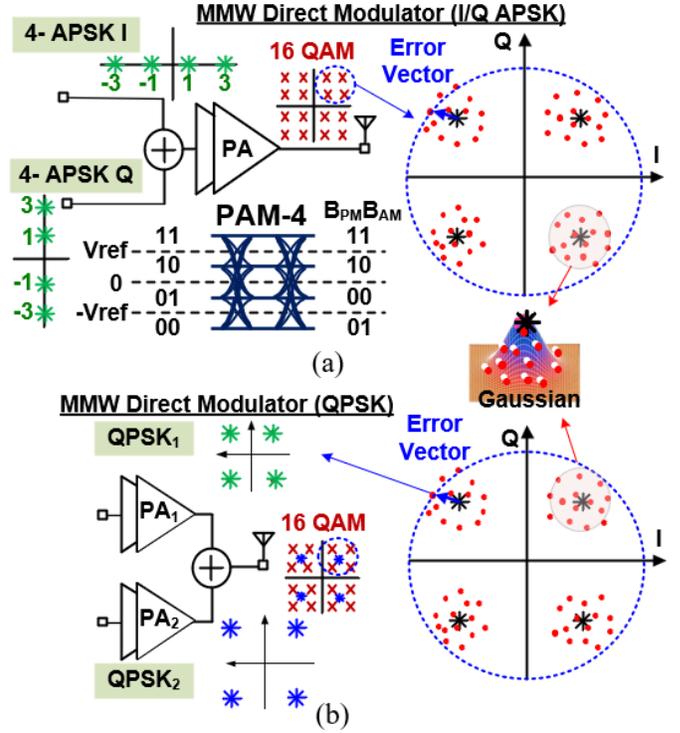

**Fig. 4:** Direct-digital MMW modulation concepts and error-vector interpretation. (a) 16-QAM generation using Cartesian I/Q modulation, [33], where each I/Q path realizes PAM-4 amplitude states (4-APSK on I and Q) and the combined I/Q vector forms the 16-QAM constellation. (b) Higher-order QAM generation by coherent vector summation of multiple QPSK paths with different amplitude weights, [34, 35], where the weighted sum produces the desired 16 QAM towards 64 QAM constellation.

### A. Qubit Operation and Driver Circuit:

In superconducting quantum systems, each qubit consists of a nonlinear LC resonator and control over qubit evolution is achieved by applying calibrated microwave pulses via a small coupling capacitor. Fig. 3. Superconducting transmon qubits typically operate at 4–7 GHz [12], and recently moving to MMW 20 GHz, [11], 70 GHz, [10], requiring on-resonance microwave pulses with controlled phase and envelope. By controlling amplitude and width of the pulse, for example $\pi$ pulse, we can transfer a single quantum of energy to the qubit. This "$\pi$ pulse" excites the qubit from its ground |0⟩ state to its excited |1⟩ state and represents the simplest quantum gate, Fig. 3(a). In quantum mechanics, the state of the qubit generally becomes a superposition of its 0 and 1 states, which can be mathematically represented by a unit vector in three-dimensional space, i.e. a point on a sphere of unit radius (so called Bloch sphere), Fig. 3(a). The power of quantum computing relies in fact in the ability to prepare qubits in such superposition.

This resonant microwave pulse applied to a qubit causes a rotation of such qubit's superposition vector on the Bloch sphere, Fig. 3(a). The rotation angle is set by the pulse area (the time-integral of the Rabi frequency, proportional to drive amplitude × pulse duration). The drive signal is modeled as a voltage waveform of the form:

$$v_q(t) = a(t)\cos(\omega_q(t) + \varphi_q), \quad (1)$$

where $a(t)$ is the envelope amplitude, $\omega_q(t)$ is the drive (carrier) frequency, and $\varphi_q$ is the carrier phase. The qubit's fundamental transition between the ground and excited states occurs at frequency $\omega_1$, and we define the detuning as $\Delta\omega = \omega_q - \omega_1$. The most common superconducting qubit, the transmon, has a second state above the 1 state that can be excited by driving the qubit at $\omega_{12} = \omega_{01} + \alpha$, where the so-called anharmonicity $\alpha \approx -(200 - 300\ MHz)$. The pulse phase $\varphi_q$ sets the rotation axis in the qubit's XY-plane, and the envelope amplitude/time integral sets the rotation angle $\theta$, [12-14]. For example, a calibrated pulse with area equal to $\pi$ (a $\pi$-pulse) will rotate the state from the north pole |0⟩ to the south pole |1⟩, Fig. 3(c), inverting the qubit population. In other words, a $\pi$-pulse (180° rotation) might be a 15 ns burst; the precise integrated area $\theta_R = \frac{g}{\hbar}\int a(t)\,dt$ determines $\theta_R = \pi$ [12]. The excited-state probability follows $P_{|1\rangle} = \sin^2(\theta/2)$, where the rotation angle $\theta$ is proportional to the pulse area (drive amplitude × duration). A half-area $\pi/2$-pulse produces a 90° rotation, taking the qubit to an equal superposition on the equator (e.g. $(|0\rangle+|1\rangle)/\sqrt{2}$).

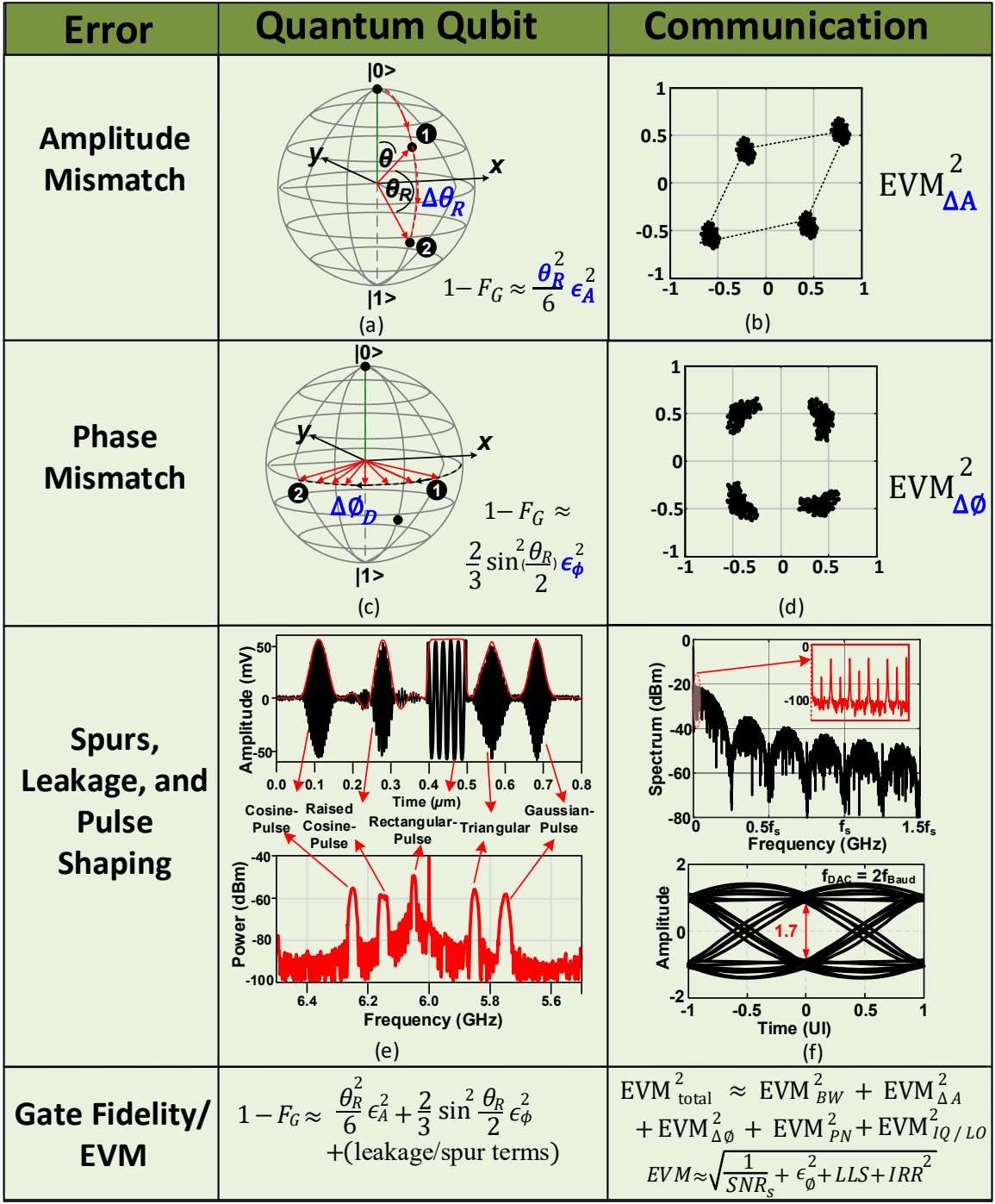

Fig. 5: Unified error interpretation for qubit control and communication using the same complex-envelope model. **Amplitude mismatch:** (a) Qubit pulse-area error produces a rotation-angle error on the Bloch sphere, contributing to gate infidelity, (b) In communications the same amplitude error moves constellation points radially and increases $EVM_{\Delta A}$. **Phase mismatch:** (c) Qubit phase error tilts the rotation axis in the XY plane and contributes to gate infidelity with $\sin^2(\theta_R/2)$ weighting; (d) In communications phase error and phase noise rotates/spreads constellation clusters and increases $EVM_{\Delta\phi}$. **Spurs, leakage, and pulse shaping:** (e) Qubit pulse shaping and spectral content illustrate leakage/spur sensitivity and the gate-time versus spectral-purity tradeoff, (f) Communication direct modulation spectrum and eye-diagram views illustrate the impact of bandwidth/settling and deterministic spurs on $EVM_{BW}$, timing margin and eye openness. (corresponding system-level metrics, relating the qubit gate-infidelity form to the communication EVM decomposition)

### B. Direct Transmitter Operation:

On the other hand, MMW communication transmitters aim to deliver complex modulation such as *M*-ASK or *M*-QAM (16-QAM or 64-QAM) directly at carriers using digitally driven RF circuits. Direct QAM generation at MMW is enabled by digital RF synthesis architectures that integrate digital control into the RF signal path, Fig. 4. A common design uses segmented transmit paths either quadrature *M*-ASK Fig. 4(a) in [33] or QPSK paths, [34], [35], Fig. 4(b).

Table I: Mapping Qubit to communication constellation

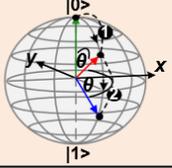

Specifically, Fig. 4(a) illustrates how baseband in-phase (I(t)) and quadrature (Q(t)) signals carry digital symbols in a 16-QAM constellation [33], [35]. In 16-QAM, each transmitted symbol corresponds to a point on a 2D plane with an I and Q coordinate. Each branch (I or Q) conveys 2 bits of information, so together they form a 4-bit symbol (16 points total). These 2 bits per branch are sent as a four-level pulse amplitude modulation signal (PAM-4) on I and on Q, [33]. In other words, each branch's amplitude takes one of four discrete levels (two bits), so the I/Q modulator maps those levels onto the X–Y constellation plans, Fig. 4(a). In another approach, in [35], several QPSK paths are coherently summed in a linear combiner (followed by the PA) instead of APSK/M-ASK, Fig. 4(b). By choosing different amplitude ratios of 1:2:4M, the system can create 16/64-QAM toward $4M$-QAM constellation based on the vector sum of weighted QPSKs, Fig. 4(b). For example, a 16-QAM symbol can be constructed by summing two QPSK signals with an amplitude ratio of 2. More generally, a $4^M$-QAM constellation is obtained recursively by adding $M$ independent QPSK paths, each scaled by successive powers of 2 in amplitude, [35].

Therefore, for direct $M$-ASK, $M$-APSK, $M$-QAM architectures, amplitude and phase shift switching and quadrature waveform generations are required. In other words, from a waveform perspective, QAM signals can be viewed as two amplitude-modulated baseband pulses in quadrature. The time-domain waveform of such amplitude modulation reveals how symbol values translate to RF amplitude variations.

C. *System* Analysis:

Per discussion in previous subsections *A* and *B*, both systems require generating an RF carrier or envelope pulse using the same complex envelope representation:

$$x(t) = I(t) + jQ(t) = A(t)e^{j\phi(t)} \qquad (2)$$

In communication (e.g., QAM), performance is determined by the envelope evaluated at symbol decision times ($t = nT_s$), where hardware imperfections appear as an error vector in the I–Q plane and are quantified by the error vector magnitude (EVM). In qubit control, the relevant quantity is not the envelope at discrete instants but its time evolution and integral over a finite gate interval of duration ($\tau_G$). Imperfections in amplitude, $A(t)$ and phase, $\phi(t)$) perturb the pulse area and effective phase, thereby changing the rotation angle and rotation axis on the Bloch sphere, which is captured by the gate infidelity $(1 - F_G)$. Therefore, the same physical impairments, finite bandwidth, amplitude/phase error, leakage, noise, and mismatch, create envelope errors that appear as EVM in communications and as gate infidelity in quantum control, enabling mitigation techniques to transfer naturally across the two domains, Table. I.

*Amplitude Error* ($\Delta A(t)$): In both systems, amplitude error originates from finite gain accuracy and nonlinearity in the envelope path (e.g., DAC code-to-amplitude nonlinearity, AM–AM/AM–PM distortion, switching-dependent impedance changes, and limited modulation depth at different amplitudes). In a communication transmitter, these errors perturb the intended constellation radius and relative spacing between points, producing a systematic error vector in the I–Q plane that raises EVM, $EVM_{\Delta A}^2$, and tightens the eye opening at high symbol rates, Fig.5(b). In a qubit controller, the same envelope-amplitude error perturbs the pulse area $\int_0^{\tau_G} a(t)\, dt$, producing a rotation-angle error $\Delta\theta_R$ that directly reduces gate fidelity, particularly for short, high-bandwidth pulses, Fig. 5(a).

*Phase and quadrature error* ($\Delta\phi(t)$): Phase error arises from voltage/digital controlled oscillator (V/DCO) phase noise, carrier phase drift, timing jitter, quadrature-generation error, and path mismatch that produces time-varying phase skew between I and Q. In communications, phase errors manifest as constellation rotation and spreading (common phase error and phase noise) and, when combined with I/Q mismatch, as image leakage and constellation distortion, Fig. 5(d), $EVM_{\Delta\phi}^2$ In qubit control, phase error perturbs the effective drive phase and therefore the rotation axis on the Bloch sphere, Fig. 5(c). This defines an average phase error during the gate as $\epsilon_\phi$ (radians), the coherent axis-error contribution is weighted by $\sin^2(\theta_R/2)$, reflecting the fact that larger-angle gates are more sensitive to axis misalignment, Fig. 5(c). Therefore, the same physical phase nonidealities that rotate and blur a QAM constellation translate into axis errors and dephasing-like control errors in qubit gates.

*Bandwidth and timing mismatch, Leakage and spurs: ISI/settling vs pulse distortion/spectral leakage*: Finite bandwidth and timing mismatch couple amplitude and phase errors by distorting the intended envelope trajectory. In communications, limited bandwidth and incomplete settling produce inter-symbol interference (ISI) and memory effects. The error vector grows with symbol rate because the envelope does not reach its intended value at decision instants, $EVM_{BW}^2$. In qubit control, finite bandwidth and non-ideal edges distort pulse shapes, redistribute spectral energy, and can increase leakage to unwanted transitions (e.g., near $\omega_{12}$). This will result in tightening the trade-off between gate time and spectral purity. Therefore, bandwidth limitations appear as ISI-driven EVM in communications and as pulse-shape and spectral-leakage-driven infidelity in qubit control, as shown in Fig. 5 (e), (f).

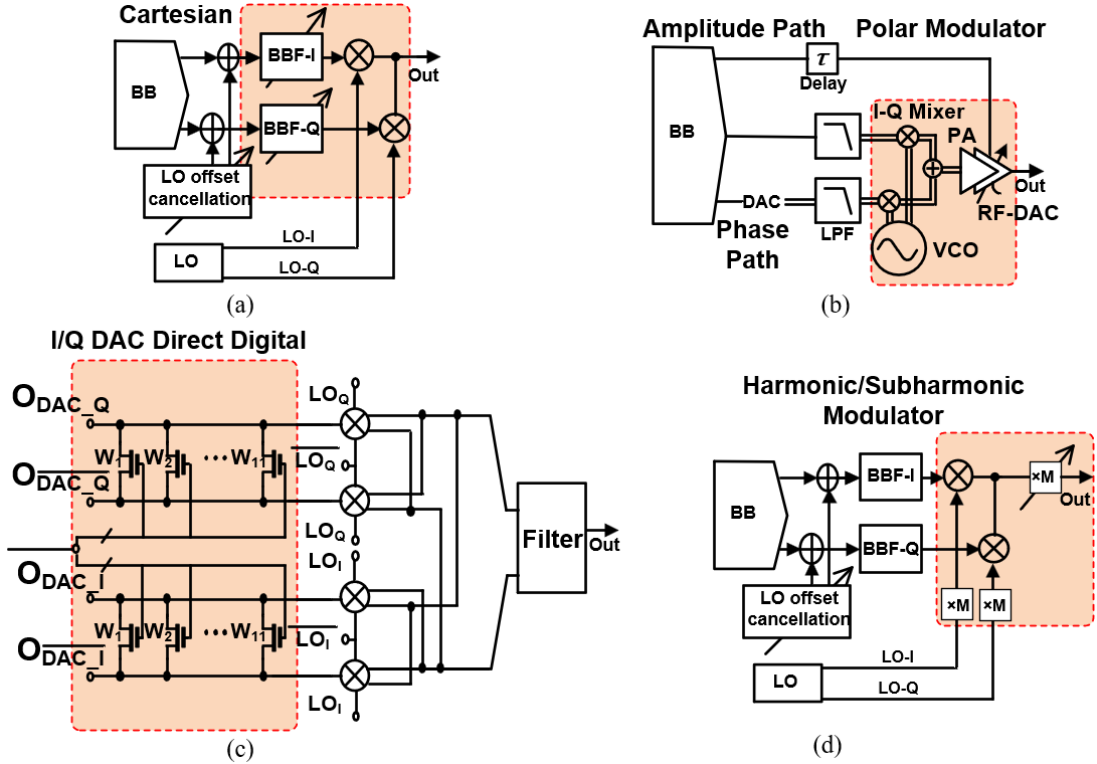

**Fig. 6:** Different circuit architecture for MMW and SubTHz communication modulator and Qubit controller. (a) IQ Cartesian Modulator/ Envelope Modulator/Mixer, (b) Polar Modulator, (c) IQ RF DAC (d) Harmonic and Subharmonic modulator (Multiplier-last).

There are other sources of leakage and spurs that represent deterministic components that are not part of the intended modulation. In direct transmitters these include LO feedthrough, DC offsets, image terms from I/Q mismatch, sampling images, and harmonic/mixing spur families. In communications they elevate in-band error and out-of-band emissions and can dominate EVM, $EVM_{IQ/LO}^2$. In qubit controllers, the analogous problem is **residual "idle" drive**: a tone near the qubit frequency during idle periods can cause unintended coherent rotations and initialization error, which drives stringent ON/OFF requirements and motivates detuning or leakage cancellation.

*Requirement of Pulsed shaping*: In both domains, reducing leakage and spur power directly reduces the deterministic component of the error vector and improves the final performance metric such as gate fidelity. One of the mitigation approaches for MMW direct modulator is pulse shaping to reduce spectral leakage, in analogy to what is performed for qubit control. The roll-off factor allows trading spectral width for timing robustness. Fig. 5(e) and (f) conceptually shows how the eye diagram evolves under different pulse shapes and with different filter roll-off factors. For example, $\beta = 1.0$ indicates that the filter's occupied bandwidth is 1.0 excess beyond the Nyquist limit (so the total bandwidth is doubled compared to an ideal Sinc pulse). Increasing $\beta$ opens the eye diagram (greater timing margin and larger vertical eye opening) at the cost of using larger bandwidth. Similarly, in qubit control there is a trade-off between pulse duration and bandwidth, where a narrowband pulse minimizes spectral leakage at the expense of gate time. In addition, special pulse shapes, such as Derivative Removal by Adiabatic Gate (DRAG), [41], achieve short duration and low gate error by suppressing specific spectral components resonant with higher transmon energy levels.

*EVM and gate-fidelity expressions:* With these sources defined, the communication error budget can be approximated and evaluated as total dependencies of these factors as:

$$EVM_{total}^2 \approx EVM_{BW}^2 + EVM_{\Delta A}^2 + EVM_{\Delta \emptyset}^2 + EVM_{PN}^2 + EVM_{IQ/LO}^2 \quad (3)$$

where each term corresponds to a specific envelope-level impairment: bandwidth/settling limits, amplitude-state errors, and leakage between levels, static or dynamic phase error, LO phase noise/jitter, and I/Q imbalance and LO feedthrough.

For qubit control, the corresponding coherent gate-infidelity contribution can estimate using the fractional pulse-area error $\epsilon_A$ and average phase error, $\epsilon_\phi$ as:

$$1 - F_G \approx \frac{\theta_R^2}{6}\epsilon_A^2 + \frac{2}{3}\sin\left(\frac{\theta_R}{2}\right)^2 \epsilon_\phi^2 + \text{(leakage/spur terms)} \quad (4)$$

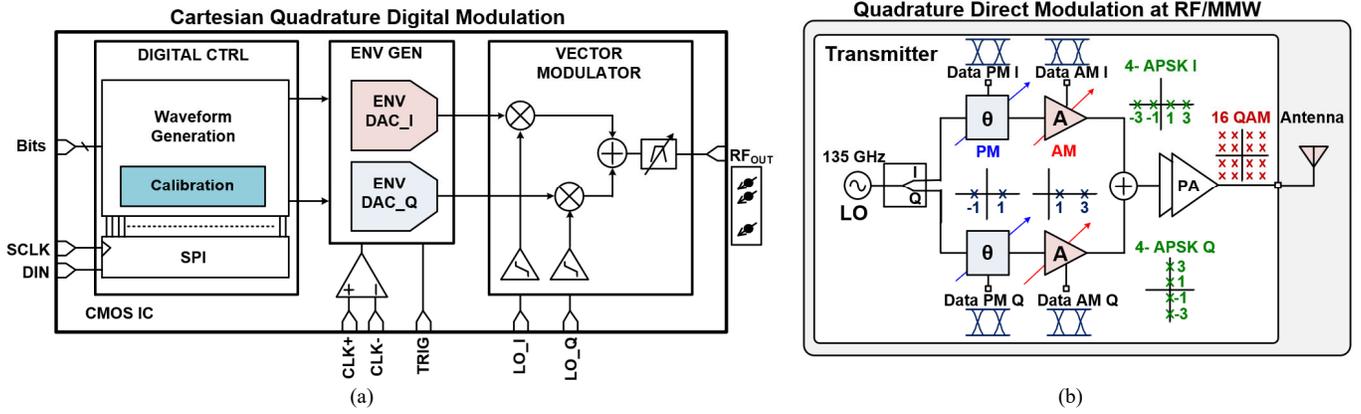

**Fig. 7**: I/Q Cartesian architecture: (a) Cartesian I/Q Envelope modulator for Qubit control [12], (b) MMW Direct modulator with amplitude and phase shift.

The derivation in (4) is consistent with the intuitive fact that large-angle rotations are most sensitive to axis misalignment, [12]. The system analysis and equations (3), (4) make the cross-domain mapping explicit as summarized in Fig. 5 and Table. I. Amplitude errors and phase errors are the dominant coherent terms in both domains, while leakage and spur terms set additional floors through unintended tones and spectral and the needs for additional filtering and pulse shaping.

Next subsection we will discuss different architecture to mitigate these nonidealities and improve the EVM and gate fidelity, $1-F_G$. *We therefore expect that advancement in one field will directly translate into the other field.*

### III. MM-WAVE DIRECT TRANSMITTER AND CRYOGENIC QUBIT GATE CONTROLLER ARCHITECTURE

Given the demonstrated synergy and coherence between the two hardware domains, Qubit and MMW communication transmitter, and their shared transformation between physical phenomena and digital information, we analyze and review the state-of-the-art architecture and hardware solutions for direct transmitter and controllers in both domains. We believe that advancements in one will directly inform and accelerate progress in the other. The main architecture for both domains consists of: *A. Cartesian IQ* mixer and vector modulator, Fig. 6(a), *B. Polar transmitter/modulator*, Fig. 6(b), *C. RFDAC transmitter* and synthesis, Fig. 6(c), and *D. harmonic/sub-harmonic* based modulator, Fig. 6(d).

#### A. Cartesian I/Q Architecture

In Cartesian I/Q modulators, two orthogonal LO phases are weighted by the baseband signals $I(t)$ and $Q(t)$ and recombined to form a complex RF output (Fig. 6(a)), and Fig. 7(a)). A single LO is split into quadrature phases (e.g., with a 90° hybrid), each branch is independently modulated, and the two paths are summed to synthesize $I(t) + jQ(t)$. In single-sideband operation, image rejection relies on tight gain and phase matching between the I and Q paths. Therefore, architecture directly targets the same envelope error terms defined in Section II, namely amplitude error $\Delta A$ and phase/axis error $\Delta \phi$. In practice, meeting low EVM in communications or high gate fidelity in qubit control requires accurate quadrature generation, low I/Q gain–phase mismatch, and strong suppression of LO leakage and DC offsets (Fig. 5).

For qubit control, the I/Q approach in [12] emphasizes envelope accuracy, spectral purity, and low power (Fig. 7(a)) [12]. The I and Q envelopes are synthesized using current-mode DAC arrays, and the stepped waveforms are low-pass filtered to produce smooth pulses. The two envelopes then pass through polarity control to realize the required phase relationship before upconversion in a passive mixer driven by a transformer-coupled LO. Since the qubit rotation angle is set by pulse area and the rotation axis is set by the drive phase, DAC linearity and settling primarily determine $\Delta \theta_R$, while quadrature accuracy and phase stability primarily determine $\Delta \phi_D$. In this architecture, residual I/Q imbalance and carrier feedthrough are especially critical because a leakage tone near the qubit band can act as an unintended "always-on" drive during idle windows. Finite baseband and mixer bandwidth also distorts pulse edges and increases spectral leakage. As a result, the implementation has optimized symmetric routings, programmable calibration of DAC weights and offsets, and leakage management. In [12], LO leakage was addressed through feed-forward cancellation in the measurement setup, highlighting the need for explicit leakage-cancellation approach in scalable implementations [12].

For MMW direct modulation, the 135-GHz direct-digital I/Q transmitter in [33] implements the same vector synthesis entirely at RF without an IF stage or broadband Nyquist DAC (Fig. 7(b)). A low-frequency LO is multiplied to the MMW band, and a coupled-line quadrature hybrid generates the I/Q carriers. The residual phase error is corrected using tunable trimming at the hybrid's isolation port. Each branch applies switch-based phase modulation (binary phase states) followed by two-level amplitude modulation. Therefore, each I and Q branch realize four APSK states (2 phase × 2 amplitude), and the recombination forms a 16-QAM constellation at the output, Fig. 4(a) and 7(b). This RF-domain approach enables high carrier operation, but it also exposes dynamic error mechanisms that directly map to $EVM_{\Delta A}$ and $EVM_{BW}$. The fast switching injects kickbacks, induces state-dependent impedance variation, and produces data-dependent settling and ISI-like distortion. In [33], a replica modulator driven with inverted data is used to cancel data-dependent kickback and reduce input-impedance variation. In addition, high-isolation combining networks (e.g., Wilkinson/transformer structures) are used to limit mutual loading so that the output combination does not convert load modulation into time-varying amplitude and phase error. In addition, per-branch gain compensation and delay matching further reduce residual I/Q mismatch and ripple-induced phase error, which would otherwise raise $EVM_{\Delta\emptyset}$ and $EVM_{IQ/LO}$ [33].

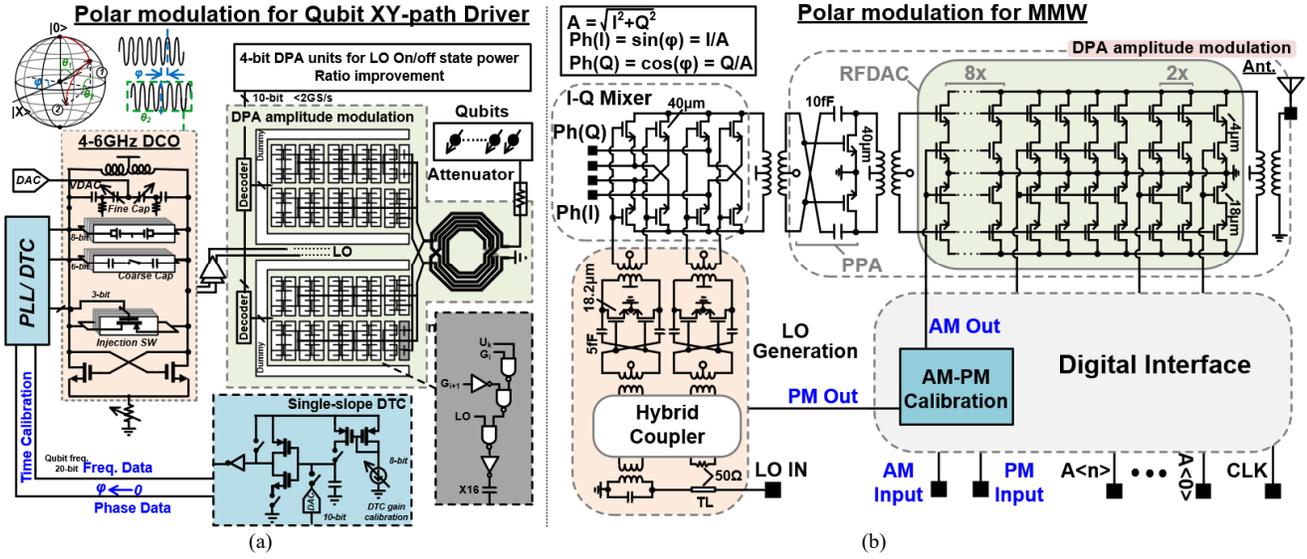

Fig. 8: Polar modulator: (a) Digital power amplifier and DCO injection locking for qubit control [16], [17], (b) MMW Direct modulator in [36] using RF DAC based digital PA.

In summary, across both qubit control and MMW communications, Cartesian I/Q provides independent control of $I(t)$ and $Q(t)$ and is therefore the most direct path to minimizing $\Delta A$ and $\Delta \phi$ in the synthesized envelope, which are the dominant drivers of both EVM and coherent gate error. Its main vulnerabilities are also shared: I/Q imbalance produces image distortion in communications and axis error in qubit rotations. The LO leakage and DC offsets create carrier feedthrough in communications and residual idle drive-in qubits; and finite bandwidth in the envelope path maps to $EVM_{BW}$ in communications and pulse distortion/leakage risk in qubit control. High-performance Cartesian implementations therefore rely on symmetric layout and accurate quadrature generation, calibration of gain/phase mismatch and leakage/offset cancellation, and transient-aware design that limits switching-induced impedance variation and prevents networks from reintroducing time-varying envelope errors.

### B. Polar Modulator

Polar modulation implements the same complex envelope $x(t) = I(t) + jQ(t)$ in polar form, $x(t) = A(t)e^{j\phi(t)}$, by separating the waveform into an amplitude path and a phase path, Fig. 6(b). The main benefit is efficiency: the phase path can be kept near constant-envelope so the power amplifier (PA) (or a digital PA (DPA)) can operate close to saturation, while the amplitude path independently sets the envelope magnitude. The corresponding cost and challenge are reconstruction sensitivity. This is mainly because the RF waveform is formed by recombining two paths, therefore, relative delay, bandwidth mismatch, and leakage in either branch directly translate into coupled $\Delta A(t)$ and $\Delta \phi(t)$ errors. These raise EVM in communications and appear as rotation-angle and rotation-axis errors in qubit control.

For qubit, the polar controller implements envelope control with a switched-capacitor digital power amplifier and phase control with an open-loop fractional injection-locked LO (ILO) driven by a digital-to-time converter (DTC) and a delta–sigma modulator (DSM), Fig. 8(a), [16]. In this polar partitioning, the amplitude path sets the pulse area, and therefore the rotation angle, while the phase path sets the carrier phase, and therefore the rotation axis on the Bloch sphere, consistent with the $(\epsilon_A, \epsilon_\phi)$ mapping in Fig. 5 and equation (4). The implementation in [16], uses a 10-bit, 2-GS/s switched-capacitor DPA and a 4–6-GHz LO path. Phase resolution is established by a single-slope 10-bit DTC with a 0.3-ps minimum step and an integrated gain-calibration range, complemented by a one-time pre-calibration to correct nonlinearity in the RF circuitry that spans from room temperature electronics to cryogenic components. The LO frequency is first set coarsely and then refined by a duty-cycled all-digital Frequency-Locked Loop (FLL) before transitioning to injection locking for stable phase during pulse delivery. To reduce average power, the DPA is duty-cycled between pulses, and an active cancellation sub-array suppresses off-state leakage and spurious output that would otherwise function as an unintended idle drive and degrade the effective on/off ratio.

For MMW transmitters, polar modulation improves efficiency by separating phase and amplitude control so the PA can operate near saturation. In the 60-GHz example, [36], the phase path is generated through I/Q up conversion, while the amplitude path modulates an RF-DAC, Fig. 8(b). The design uses 10-GS/s sampling (6× oversampling) to suppress aliasing and support the required baseband bandwidth. However, the achievable EVM is still limited by amplitude–phase time alignment, RF-DAC nonlinearity, and residual leakage at small amplitude codes. The polar transmitter shows that minimum EVM requires explicit delay tuning between the amplitude and phase paths; for example, with synchronization using a 1.5-ns delay with 50-ps resolution. At output power back-off, the EVM floor is dominated by system nonidealities, primarily residual leakage and amplitude-dependent gain distortion (AM–AM) and amplitude-to-phase conversion (AM–PM) in the RF-DAC/PA chain. This motivates the digital predistortion and amplitude-path calibration.

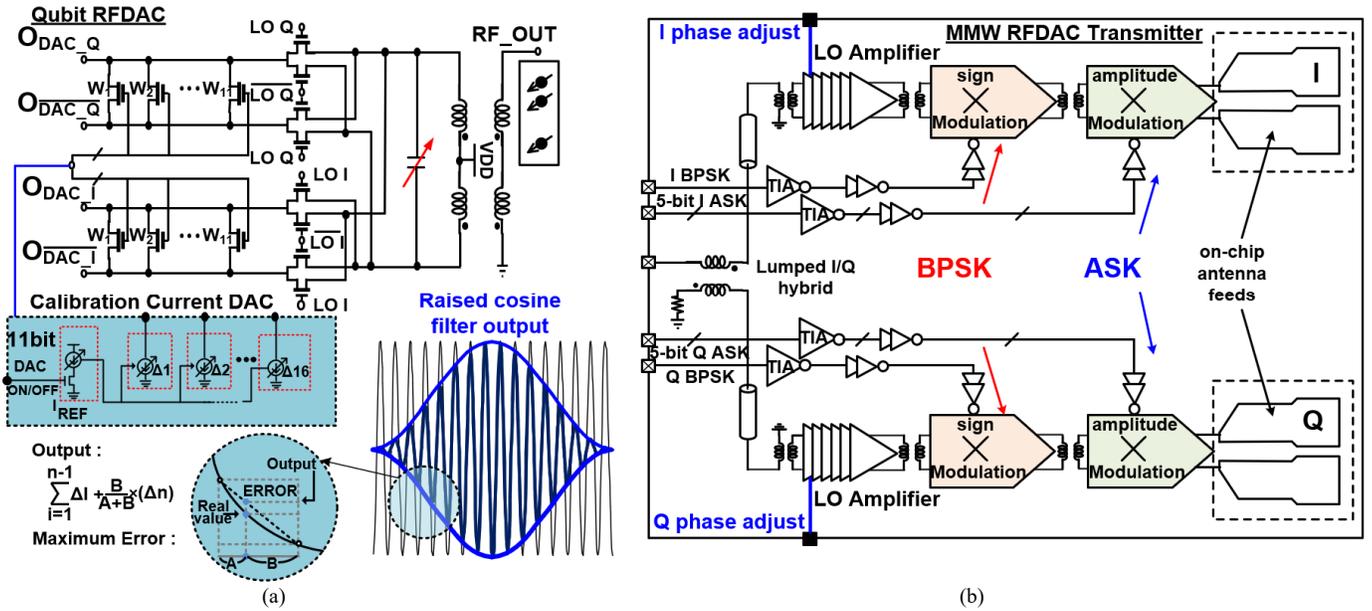

Fig. 9: Direct I/Q RF DAC transmitter, (a) Direct I/Q RFDAC with Raised Cosine Filter and 11-bit current steering DAC for calibration, [18], (b) MMW I/Q RF power-DAC including ASK and BPSK modulators in I and Q path, [38], generating up to 64QAM.

Across both domains, the dominant polar-design constraint is therefore reconstruction fidelity. Delay mismatch and bandwidth mismatch between amplitude and phase paths appear as a coupled error vector, simultaneously creating effective $\Delta A(t)$ and $\Delta \phi(t)$. In communications this raises EVM; in qubit control it appears as combined $\Delta \theta_R$ and $\Delta \phi_D$, together with sensitivity to residual off-state leakage. As a result, practical polar systems rely on explicit timing alignment (programmable delay/retiming), careful bandwidth partitioning and filtering in the two paths, and calibration/predistortion to manage AM–AM/AM–PM distortion and suppress leakage, particularly at small amplitudes and during nominal "off" intervals. In summary, polar modulation is attractive for saturated MMW transmitters and low-power cryogenic XY drivers, but its performance is mainly determined by amplitude–phase alignment, path matching, and leakage/spur suppression rather than by the polar transform itself.

### C. I/Q RF DAC

The direct I/Q RF-DAC (or DDS-assisted RF-DAC) architecture is attractive in both cryogenic qubit control and MMW transmitters because it synthesizes the desired complex envelope digitally and applies it to an RF modulator with a compact DAC/mixer front end, enabling amplitude- and phase-modulated carriers directly at, or near, the RF frequency, Fig. 6(c). Its key advantage is that waveform shaping is performed at the baseband. The envelope can be programmed with fine tunability and its spectrum controlled through reconstruction filtering and pulse shaping. This will directly address the bandwidth and spectral-leakage limitations discussed in Section II, Fig. 5(e) and (f). This approach does not eliminate the LO/carrier path; instead, it reduces dependence on broadband analog baseband chains and makes calibration and spectral shaping embedded for the transmitter/controller.

For superconducting platforms, the DDS implementation has been reposted in [19], [18], [2]. A representative cryogenic implementation in [18] consists of independent pulse modulators operating at a shared LO frequency, shown in Fig. 9 (a), [18]. Each modulator uses a two-stage DAC topology. A nonlinear sinusoidal-shaping DAC generates coarse waveform structure from 4 most significant bit (MSB) control bits, while a fine-grained linear interpolating DAC handles the 4 least significant bit (LSBs) to achieve smooth transitions and higher resolution. The combined DAC output is mixed with quadrature LO phases to produce a spectrally isolated RF signal. The raised cosine filters are included to suppress spectral sidelobes, implemented through a shunt-based shaping network that approximates a high roll-off factor. To compensate for process variation and ensure signal symmetry across qubits, each DAC includes on-chip calibration current sources with ±20% range and 11-bit resolution as shown in Fig. 9(a). Simulation results indicate that this hybrid nonlinear-linear DAC achieves an effective resolution of approximately 9 bits. This enables qubit pulses with accurate phase and amplitude modulation, tunable duration (10–100 ns typical), and burst-mode operation, all under strict power budgets in the range of 2.9 mW for analog and digital part.

In parallel, direct I/Q DAC architectures are deployed in MMW and sub-THz systems to support high-order modulation while managing the loss and isolation challenges of on-chip combining at >100 GHz (Fig. 9(b)). In [38], instead of summing I and Q on-chip using a passive coupler, the I and Q signals are radiated from spatially separated antennas and recombined over the air. This free-space quadrature summation avoids the loss and finite isolation of dense RF combiners, mitigates load pulling between branches, and benefits from spatial array gain. The transmitter uses a two-stage modulation structure: a BPSK (sign) modulator in the early stage sets the carrier phase, followed by a segmented ASK stage near the output to set the carrier amplitude (Fig. 9(b)). However, gate segmentation increases modulation depth while reducing capacitive loading, which is critical for constellation accuracy at D-band. However, the achievable minimum amplitude and modulation depth remain limited by parasitic capacitances and device leakage at high frequency. Therefore, low-amplitude constellation points are particularly sensitive to residual leakage and code-dependent distortion. Accurate differential switching, proper segmentation, and phase alignment are essential to suppress feedthrough and preserve signal quality at small symbol amplitudes, while programmable phase shifters enable real-time steering of the information beam.

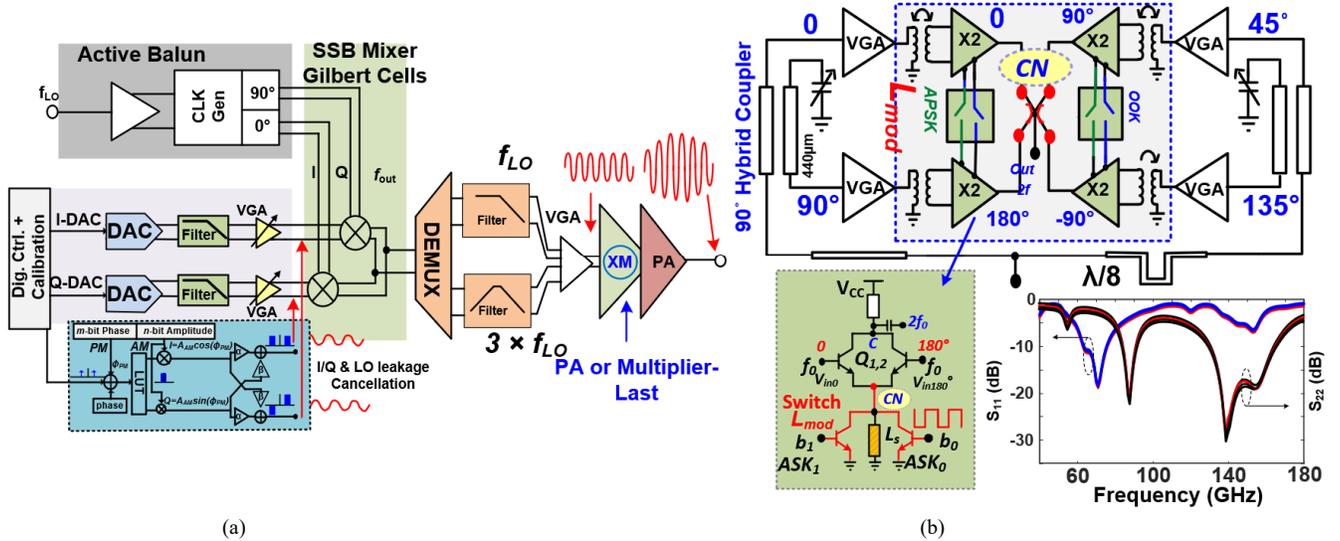

**Fig. 10:** Harmonic and subharmonic transmitter: (a) modified Qubit controller with dual band controller, [20] and [15], (b) Northeastern's harmonic transmitter with OOK and APSK modulator for sub-THz band direct modulation with constant impedance [26].

Across both domains, the practical limitation of RF-DAC/DDS is not the concept of digitally programming the waveform, but how accurately the DAC-based RF modulator can realize small amplitude and phase steps while keeping the spectrum clean. The key challenges therefore lie in spectral shaping and reconstruction, calibration under mismatch and drift, modulation linearity, and managing leakage caused by layout parasitics. In addition, the effective number of bits (ENOB) and clock/jitter limits that set the spur and noise floor should be considered. As a result, high-performance RF-DAC implementations in both domains require a combination of calibration (to correct static mismatch and restore symmetry), spectral control (reconstruction filtering and pulse shaping to suppress images and sidelobes), and careful switching/segmentation strategies to minimize leakage and code-dependent distortion.

In summary, RF-DAC/DDS provides a common digital-to-RF synthesis path for both qubit control and MMW communication. In cryogenic qubit controllers, it enables programmable pulse generation with built-in shaping and calibration under tight thermal and power constraints. In MMW transmitters, it supports direct high-order modulation. In both domains, performance is ultimately limited by spectral purity and signal accuracy, specifically modulation depth at low amplitude settings, code-dependent spurs and leakage, and ENOB/clock-jitter constraints. Therefore, practical designs rely on segmentation, calibration, and spectral shaping to prevent these effects from setting the EVM or gate-error floor.

*D. I/Q Harmonic/Subharmonic Modulator*

Harmonic and subharmonic modulators relax LO generation and distribution by deriving the RF carrier from a lower-frequency reference. In harmonic operation, the carrier is produced by frequency multiplication (e.g., Doubler/Tripler chains) and modulation is applied at the harmonic, Fig. 6(d). In subharmonic mixing, the LO runs at a fraction of the target RF and mixer nonlinearity generates the desired band (Fig. 6(d)). This is useful in both qubit controllers and MMW/sub-THz arrays, where distributing a high-frequency, low-phase-noise LO to multiple channels is difficult. Using a lower-frequency reference reduces the number of high-frequency distribution paths and simplifies frequency planning at scale.

A representative cryogenic example is the controller in [20], which covers 2–20 GHz from a single external LO by splitting the RF path into a 2–15 GHz low band mixed at the fundamental and a 15–20 GHz high band generated by steering mixer current into a tank tuned near $3f_{LO}$ (Fig. 10(a)). The $3f_{LO}$ branch extends tuning to 20 GHz without a second synthesizer and preserves the same I/Q chain and calibration knobs used in the low band. The tradeoff is additional spur content and reduced conversion efficiency. Therefore, the design relies on I/Q and offset calibration and reconstruction filtering to suppress image and carrier terms and maintain high Spurious-Free Dynamic Range (SFDR) in the qubit band.

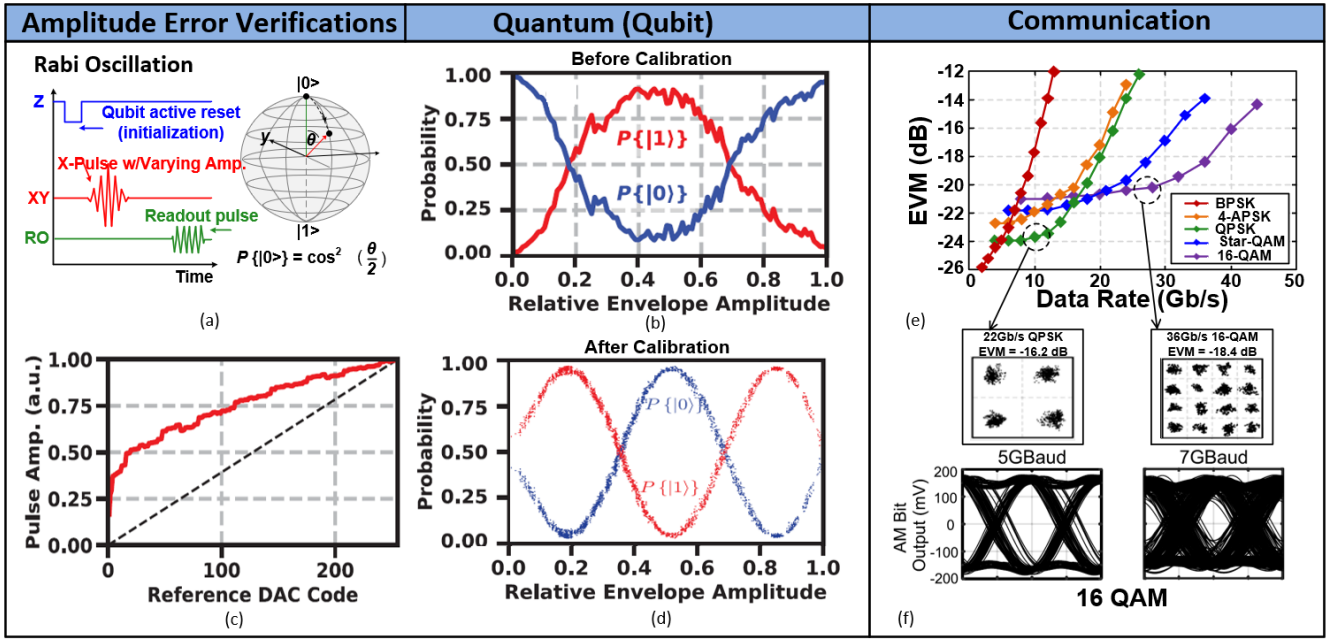

**Fig. 11: Amplitude-error verification (qubit and communication):** (a) Single-pulse Rabi protocol illustrating initialization, XY drive with amplitude sweep, and readout in [12], (b) Uncalibrated Rabi response $P\{|0\rangle\}$, $P\{|1\rangle\}$ versus relative envelope amplitude, (c) Measured envelope amplitude versus reference DAC code demonstrating nonlinear/non-monotonic code-to-amplitude behavior prior to calibration, (d) Calibrated Rabi oscillations versus envelope amplitude confirming restored pulse-area control and rotation-angle accuracy; (e) Measured EVM versus data rate for a 135-GHz direct-digital link across modulation in [33], showing EVM degradation with higher symbol rate and modulation order. (f) Example 16-QAM demodulated eye diagrams at two baud rates, illustrating eye closure at higher rate due to bandwidth/settling limits.

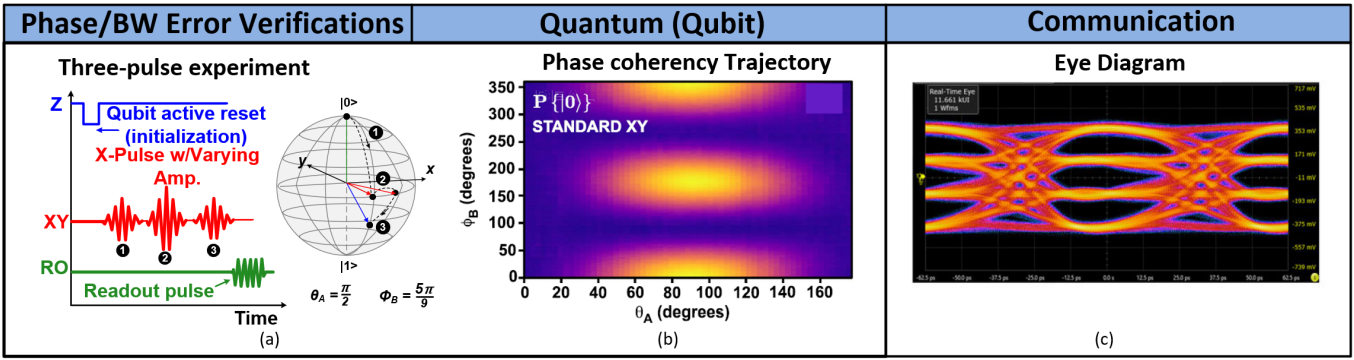

**Fig. 12. Phase and bandwidth-error verification (qubit and communication).** (a) Three-pulse phase-coherency experiment in [12]: a sequence of XY pulses with swept amplitude and a controlled phase step, followed by readout, (b) Measured phase-coherency map $P\{|0\rangle\}$ versus $\theta_A$ (amplitude/rotation) and $\phi_B$ (phase step) from [12], highlighting deterministic axis/phase errors observable under fast switching. (c) Communication eye diagram showing timing margin and ISI/settling effects; eye closure provides the comm-domain analogue of amplitude/phase/BW sensitivity.

At MMW/sub-THz, harmonic/subharmonic modulation is often used in a multiplier-last approach to avoid distributing a high-frequency LO across multiple elements. These schemes can support high-speed OOK/ASK/PSK, with quadrature paths, and higher-order modulation by embedding switching and modulation around the multiplier [39], [26]. The practical limitation is switching behavior at very high speed. Overshoot, ringing, and finite settling translate into memory, eye closure, and higher EVM, while impedance variation, such as Voltage Standing Wave Ratio (VSWR) mismatch between PA and antenna or between LO distribution and switching networks, creates element-to-element amplitude and phase error that limits coherent combining.

Conventional amplitude switching, such as PA-bias OOK [33] or conversion-gain tuning in mixers [38], often suffers from load modulation and rise/fall asymmetry, which increases spectral leakage. The recent proposed harmonic On-Off Keying (HOOK) architecture [24]–[26] reduces these effects by placing the modulator at a virtual-ground common node of a differential push–push doubler (Fig. 10(b)). In this configuration, the harmonic current is modulated while the fundamental sees a near-constant impedance boundary. This limits state-dependent impedance variation during switching and improves transient symmetry. This constant-impedance behavior also helps array scaling by reducing switching-induced loading differences across elements.

The main challenge of harmonic/subharmonic translation is spur and phase-noise control. Multiplication and harmonic extraction introduce additional spur families that must be filtered or canceled, and LO phase perturbations become more critical after multiplication (phase-noise power increases roughly by $20\log_{10} M$). Gain ripple and impedance variation can further introduce $\text{EVM}^2_{\Delta A}$ and amplitude-to-phase conversion that appear as envelope error over bandwidth, $\text{EVM}^2_{BW}$. Practical designs therefore rely on filtering, symmetry in switching/layout, and calibration of leakage and image terms. In summary, harmonic and subharmonic modulators are well suited for scalable and wideband LO planning, but performance is set by spur management, phase-noise budget, and conversion ripple control.

## IV. STATE OF THE ART MEASUREMENT RESULTS AND COMPARISONS RESULTS

This section presents the representative measurement results for Cartesian I/Q, polar, RF-DAC, and harmonic/subharmonic architectures and shows how calibration and tuning reduce the dominant nonidealities introduced in Section II and III. For each architecture, we emphasize the error term that most clearly limits performance and use the corresponding measurement observable. Pulse-area error is evaluated using Rabi amplitude mapping ($\Delta A$, $\Delta\theta_R$), Fig. 11, while phase and timing errors ($\Delta\phi$) are evaluated using phase-coherent tests and EVM/constellation (eye diagram) behavior, Fig. 12. Bandwidth and settling limits are captured through EVM versus data rate and the measured SSB response ($\text{EVM}_{BW}$). Deterministic tones and spur control are assessed using spectrum/SFDR and on/off leakage measurements, (Fig. 13, Fig. 14), with calibration implemented through AM–PM correction or DPD on the communication side and pulse shaping on the qubit side.

*A. Cartesian Transmitter (Amplitude and Phase Error Test):* For Cartesian I/Q, the main performance analysis and measurement verification is accurate vector synthesis. Gain and phase mismatch between the I and Q paths appear as image leakage and constellation rotation in MMW links, deteriorating EVM and the constellation diagram, and the same mechanism appears as a rotation-axis error in qubit control (Fig. 5(a), (b)). In addition, envelope nonlinearity and finite settling contribute to $\Delta A(t)$ and $\text{EVM}_{BW}$, and LO/DC leakage contributes to $\text{EVM}_{IQ/LO}$. These terms can corrupt both constellation quality in communications and pulse fidelity in qubit control.

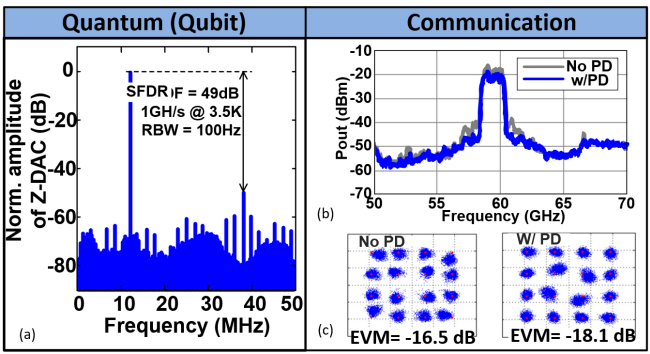

**Fig. 13:** Polar-modulation measurement highlights (qubit and communication). (a) Measured SFDR of the qubit-controller DAC in [16] (b) Measured MMW output PSD with and without predistortion in [36], , (c) Measured 16-QAM constellations for the 60-GHz polar transmitter without predistortion and with predistortion.

The first validation step is amplitude linearity, because the pulse area sets $\theta_R$. Fig. 11(a) shows the Rabi oscillation protocol and the Bloch-sphere trajectory for the single-pulse experiment. The uncalibrated amplitude performance of the I/Q controller is shown in Fig. 11(b), where the measured envelope amplitude does not follow a linear mapping with the programmed reference DAC setting. As a result, the pulse area does not scale linearly with the programmed code, and the Rabi response shows systematic over- and under-rotation. In [12], this behavior is traced to non-monotonicity and non-linearity in reference DAC during amplitude sweeps, which is captured in Fig. 11(c). After envelope calibration, Fig. 11(d) shows high-visibility Rabi oscillations with the expected sinusoidal dependence on the calibrated envelope amplitude, confirming that the controller can reliably program pulse area and therefore $\theta_R$ after proper DAC current calibration and weighting. This qubit-side result directly verifies the Section II mapping that amplitude error ($\Delta A$) translates into rotation-angle error ($\Delta\theta_R$).

For the MMW direct-digital transmitter, the same limitations appear in the constellation domain. Fig. 11(e) shows that the measured EVM versus data rate increases with symbol rate and modulation order from [33], consistent with finite settling, residual state-dependent distortion, and increasing sensitivity of dense constellations to $\Delta A$ and $\Delta\phi$ errors. The demodulated eye diagrams in Fig. 11(f) provide the corresponding time-domain view. As baud rate increases, eye opening reduces, indicating smaller timing and amplitude margin due to bandwidth and settling limits. In [33], the reported link performance reaches approximately $-18.4$ dB EVM for 36 Gb/s 16-QAM and $-16.2$ dB for 22 Gb/s QPSK, and the degradation at higher rates is consistent with increased sensitivity to amplitude/phase errors and load-modulation effects during fast switching. Overall, Cartesian I/Q achieves strong performance once the envelope mapping and I/Q balance are calibrated. However, the highest rates remain limited by residual I/Q mismatch and LO leakage, and, specifically for direct-digital modulation at MMW, by switching transients and bandwidth ripple that set the practical $\text{EVM}_{BW}$ and residual vector-error floor.

After amplitude linearity is established, the next verification step is phase coherency under fast switching for both qubit controller and MMW transmitter. Since the carrier phase sets the rotation axis and any phase error appears directly as $\Delta\phi_D$. Fig. 12(a) shows the three-pulse protocol used in [12] to probe this axis control. The sequence applies to an active reset, then an $X-$pulse with programmable amplitude, followed by a $\pi-$pulse whose carrier phase is intentionally stepped, and finally a readout pulse. By sweeping the first-pulse rotation angle $\theta_A$ and the phase step $\phi_B$, the experiment forces the controller to produce phase-accurate transitions over a short time window. Therefore, any carrier-phase error, quadrature imbalance, or fast-switching transient that perturbs the effective drive phase appears as a distortion in the final state probability map.

Fig. 12(b) reports the measured phase-coherence trajectory $P\{|0\rangle\}$ as a function of $\theta_A$ and $\phi_B$ for the standard XY controller. In an ideal phase-coherent controller, the probability map exhibits the expected structured dependence on the commanded amplitude and phase settings. This is consistent with the Section II model: amplitude mapping errors contribute to $\Delta\theta_R$, while phase-path errors and timing skew contribute to $\Delta\phi_D$. In [12], the fact that the error is structured motivates explicit calibration of amplitude and phase settings. Once the amplitude mapping is corrected (as verified by the Rabi calibration in Fig. 11), the three-pulse map becomes a direct way to isolate residual phase/axis error and fast-switching artifacts that are not visible in a single-pulse experiment (Fig. 12).

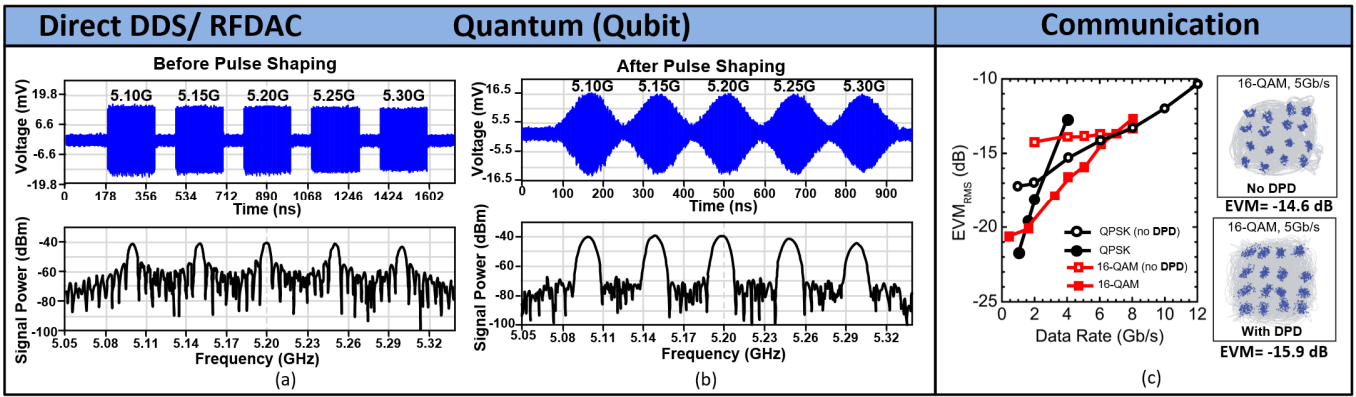

**Fig. 14: Direct RF-DAC/DDS measurement highlights (qubit vs communication).** Direct RF-DAC controller spectrum with rectangular and raised-cosine filtering from [18], measured output of the cryogenic RF-DAC qubit controller. (a) Time- and frequency-domain response for rectangular pulse modulation across 5.1–5.3 GHz. (b) Corresponding raised-cosine pulse modulation showing a smoother envelope and reduced spectral sidelobes. **(c)** MMW RF-DAC transmitter performance from [38]: measured $EVM_{RMS}$ versus data rate for QPSK and 16-QAM with and without digital predistortion (DPD).

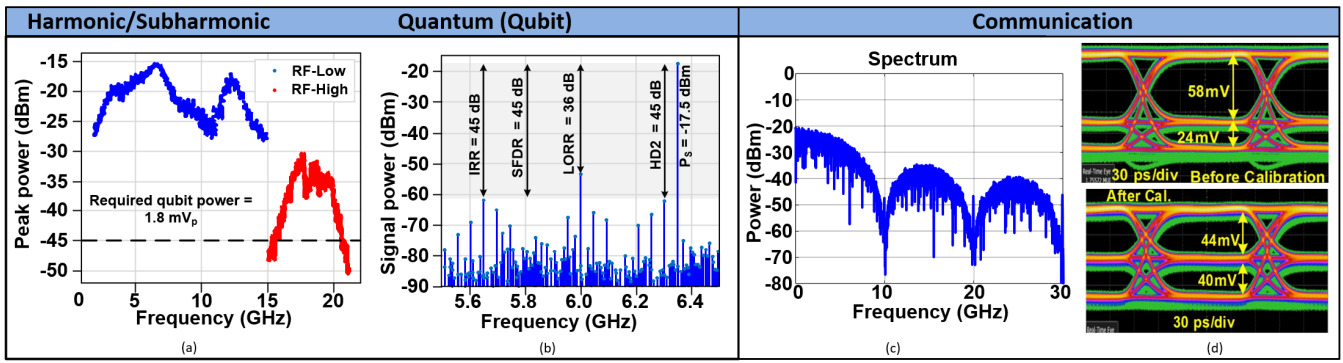

**Fig. 15: Harmonic/subharmonic measurement highlights (qubit vs communications).** (a) Measured dual-band output coverage of the harmonic/subharmonic cryogenic controller in [20], showing RF-Low and RF-High bands (2–20 GHz), (b) Spectrum of the upconverted cryogenic output illustrating spur suppression after calibration, including SFDR (~45 dB) and key rejection metrics (e.g., IRR/LODR/HD2). (c) Northeastern harmonic (HOOK) sub-THz transmitter [26]: measured high-speed OOK/ASK spectrum under direct harmonic switching, (d) Eye diagram of 3-ASK before and after calibration, showing improved level uniformity and eye opening after amplitude equalization and phase tuning [26].

The communication analogue of this test is the eye diagram in Fig. 12(c). In a direct-modulation link, the eye diagram overlays many symbol transitions, so it exposes the same class of impairments: finite settling, timing skew, and memory in the waveform that prevent the signal from reaching its intended value at decision instants. When the eye closes, the receiver sees increased decision error and a higher EVM floor, even if the average constellation appears reasonable. The three-pulse qubit experiment plays the same role. It overlays fast switching in amplitude and phase and reveals whether the controller reaches the intended rotation axis and angle at the required times. In this sense, the qubit "phase-coherency trajectory" and the communication eye diagram are parallel diagnostics of the same envelope-synthesis quality, one measured as phase-coherent state probability under multi-pulse control, and the other measured as timing and amplitude margin under repeated symbol transitions.

***B. Polar Transmitter and AM-PM Verifications***: Polar modulation separates the waveform into amplitude and phase paths, so calibration is governed by reconstruction fidelity. Any relative timing or bandwidth mismatch between the two paths appears as a coupled $\Delta A(t)$ and $\Delta\phi(t)$ error, which increases EVM in communications and contributes to rotation-angle and rotation-axis errors in qubit control. On the qubit side, the most direct calibration outcomes are spur and leakage suppression. Fig. 13(a) reports the measured SFDR of the qubit controller's DAC at 3.5 K, showing a clean spectral floor (SFDR = 49 dB at 1 GS/s), which reduces deterministic spurs that can drive unwanted transitions, [16]. The same qubit polar-controller work reports in [16] XY-path operation over the 4–6 GHz band with 13.7 mW/qubit active power, output power of at least −15 dBm across 4.1–5.96 GHz with SFDR better than 40 dB for the modulated drive, and an on/off power ratio of 38.2 dB achieved using active cancellation. these metrics map directly to the leakage/spur terms in the gate-error budget.

On the MMW side, the same dependence on reconstruction fidelity appears in spectrum and constellation quality. Fig. 13(b) shows the measured output PSD with and without digital predistortion (AM–AM/AM–PM DPD calibration), where digital predistortion reduces spectral regrowth. Fig. 13(c) shows the measured 16-QAM constellations of [36] and the associated EVM improvement from −16.5 dB (no digital predistortion DPD) to −18.1 dB (with digital predistortion DPD). The MMW polar transmitter measurements in [36] also show that minimum EVM requires explicit amplitude–phase alignment; synchronization is achieved using a 1.5-ns delay with 50-ps resolution between the amplitude and phase paths, and the reported QPSK EVM improves from −20.7 dB to −23.6 dB with predistortion. Overall, these results support the Section II conclusion for polar systems that the efficiency is high, but final performance is set by AM–PM alignment and leakage/nonlinearity control in the amplitude path.

***C. RF-DAC and Pulse Shaping:*** The RF-DAC/DDS places waveform synthesis at the BB. This makes pulse shaping and spectral control straightforward but shifts the limiting factors toward code-dependent distortion, sampling images, and signal leakage. The calibration objective is therefore twofold: enforce symmetry/linearity across DAC elements and suppress deterministic spurs and images.

On the qubit side, the RF-DAC measurements show that shaped pulses provide the intended spectral benefit. The rectangular versus shaped pulse spectra demonstrates reduced sidelobes with raised-cosine (or similar) shaping, which directly reduces leakage risk to nearby transitions. The reported IRR/LORR and spectral plots provide a practical validation that image and carrier terms are sufficiently suppressed after calibration and trimming. Fig. 14(a)-(b) shows the measured benefit of pulse shaping in the direct-synthesis controller of [18]. Without shaping, the rectangular bursts exhibit higher spectral sidelobes. By applying raised-cosine shaping, the sidelobes are suppressed and the spectrum becomes more selective across the programmed 5.1–5.3-GHz tones. This directly targets leakage cancellation in dense qubit systems by reducing spectral spurs outside the intended band. The same work reports strong image and LO rejection (IRR > 42 dB and LORR > 41 dB at 1 GS/s) and uses per-cell calibration currents (±20% range with 11-bit resolution) to correct mismatch and preserve channel symmetry.

On the MMW side, the D-band RF-DAC transmitter results highlight a complementary limitation. Even with high-order modulation capability, the EVM floor is often set by modulation depth at low amplitude codes and by parasitic leakage that distorts the inner constellation points. The constellation plots before and after equalization show that equalization can reduce some bandwidth-induced distortion, but low-code leakage and code-dependent error remain the dominant limiter for dense constellations. This matches the Section II framing: $EVM_{\Delta A}$ and leakage terms dominate at small amplitudes, while $EVM_{BW}$ can be partially corrected by equalization. The direct RF-DAC transmitter in [38] supports up to 64-QAM, with measured EVM of −19.7 dB at 3.6 Gb/s for 64-QAM. As shown in Fig. 14 (c) the EVM versus data-rate plot, the architecture maintains <−20 dB EVM up to 4 Gb/s for QPSK and below −17 dB for 16-QAM, demonstrating robust back-off performance without DPD. With DPD the EVM can be enhanced, in the reported results in [38], illustrated in Fig.14(c), 16-QAM improves from −14.6 dB EVM (no DPD) to −15.9 dB EVM (with DPD).

***D. Harmonic/sub-harmonic transmitter and spectral purity (Spurs):*** Harmonic/subharmonic schemes expand frequency coverage and relax LO distribution, but they also introduce additional spur families that must be managed. In qubit controllers, the most direct validation is spectral purity across bands. For cryogenic qubit control, harmonic/subharmonic architectures appear in two versions (Fig. 6(d)). Fig. 15(a) shows the measured RF output coverage of the dual-band cryogenic controller in [20], where the RF path is split into RF-Low (2–15 GHz) and RF-High (15–20 GHz) using third-harmonic up conversion, enabling wide coverage from a single external LO. The corresponding spur/SFDR plot in Fig. 15(b) shows that, after digital I/Q and offset calibration, image and LO leakage are controlled and the peak-output SFDR exceeds 45 dB; the reported SFDR is primarily limited by HD2 and image rejection, while LO leakage after calibration does not set the SFDR limit. This architecture reduces the number of LOs and distribution paths per qubit and relaxes frequency planning for dense frequency-division multiple access (FDMA), but it requires explicit spur planning and filtering because harmonic translation introduces additional spurs.

On the MMW side, the harmonic modulator measurements (HOOK [26]) provide a clear transient- and spur-calibration example. Fig. 15(c) shows the measured spectrum under direct OOK keying at high data rate (e.g., 30 Gb/s), where spectral regrowth and discrete spurs are set by switching edge quality and leakage inherent for direct modulators. Fig. 15(d) then provides the time-domain verifications, the eye diagram before calibration shows unequal levels and reduced opening, while after calibration, using amplitude equalization and phase tuning, the eye opens and the levels become more uniform. This improvement is consistent with the core HOOK advantage: constant-impedance switching reduces switching-dependent loading, so calibration can focus on layout-driven amplitude and phase errors rather than compensating large state-dependent impedance swings. The quadrature second-harmonic modulator achieves high OOK depth, with an on–off switching isolation of approximately 25 dB, and a 2-ASK depth of 6–8 dB over a measured 20% fractional bandwidth (132–158 GHz). The system supports up to 45 Gb/s at 3.3 pJ/bit with the measured EVM is better than −17 to −19 dB.

## V. COMPARISON AND FUTURE DIRECTIONS

Among the four direct modulation strategies evaluated, Cartesian I/Q, [12-15], [33], Polar, [16], [17], [36], RF-DAC, [18], [19], [38], and harmonic/subharmonic, [20], [24-26], each offers distinct trade-offs in energy efficiency, calibrations, spectral purity, and scalability. Table II and III summarize the comparison of these architectures for both qubit control and MMW transmitters, including recent state-of-the-art implementations.

Tabel II: Comparison between different Direct MMW Transmitter architecture

| Architecture | Cartesian I/Q [33] | Polar [36] | Direct I/Q RF-DAC [38] | Harmonic / Sub-harmonic [26] |
|---|---|---|---|---|
| Frequency (GHz) / Techlonogy | 135 / 28 nm CMOS | 60 / 65 nm CMOS | 140 / 45 nm CMOS | 140/ 90 nm SiGe |
| Modulation Scheme | 16-QAM / QPSK | 16-QAM / QPSK | 64-QAM / QPSK | Star-QAM / OOK |
| Measured Peak Data Rate (Gb/s) | 36 (16-QAM) | ~10 | 3.6 (64-QAM) 12 (QPSK) | 30/45 (Star-QAM) |
| EVM (dB) | -18.4 (16-QAM) -16.2 (QPSK) | -16.5 (16-QAM) -20.7 (QPSK) | -19.7 (64-QAM) | -17 (Star-QAM) |
| Output Power (dBm) / Bit Efficiency (pJ/bit) | 0 / 9 | 10.8 / NA | 13.2$^{*EIRP}$ / NA | -3.6 / 2.2-3.3 |
| Key Circuit Features | Switch-based PM +2-level AM | Digital PA/Phase Modulate | Spatial I/Q Combining + Phase-offset Correction | Double HOOK + Constant Impedance |

Tabel III: Comparison between different Qubit Gate Controller

| Architecture | Cartesian I/Q [12] | Polar [16] | Direct I/Q RF-DAC [18] | Harmonic / Sub-harmonic [20] |
|---|---|---|---|---|
| RF Band (GHz) / Clocks (GS/s) | 4-8 / 1 (AWG or DACs) | 4-6 / 2 (DPA) | 2-7 / 1-2 (DDS) | 2-20 / NA |
| Spectral Metrics | Low-spur LO- Shaped Pulses | SFDR > 40dB ON/OFF ratio = 38dB | IRR > 42 dB LORR >41 dB | SFDR ≈ 45dB IM3 ≈ 50dB |
| Active Power (mW) per Qubit / Channel | < 2 | 13.7 | ≈ 5.5 | ~ 1.7 (Analog) 330 (Digital) |
| Validation / Capability | Coherent Rabi Gate Demos with IC | Arbitrary XY Pulses, Rabi Oscillation Demo | Rabi/Ramsey (77K) | Rabi Shown at High GHz Bands |
| Main Feature | Low-Power + Fixed LO | High efficiency + Need A/φ Alignment and Spur Control | Full freq./phase/amp. Programmability | Wide RF Range + More Spurs to Calibrate |

Cartesian I/Q is the most direct path to envelop accuracy because it synthesizes $I(t)$ and $Q(t)$ explicitly. When gain/phase and offset are calibrated, it can suppress the dominant coherent errors in both domains: amplitude/area error that sets $\Delta\theta_R$ in qubits and radial constellation error in communications, and phase/axis error that sets $\Delta\phi_D$ in qubits and constellation rotation in communications. Its main vulnerability is also shared between two domains. The small I/Q mismatch produces an image term, and LO feedthrough/DC offsets create a residual carrier. In qubit drives this residual tone can behave like an unintended "idle" excitation if not canceled or detuned, while in MMW links it appears as carrier leakage and additional EVM and mask pressure. As symbol rates increase, finite bandwidth and switching transients become the practical limiter because they introduce memory and settling distortion that cannot be removed by static calibration alone.

Polar modulation improves efficiency by separating amplitude and phase control so the power stage can operate near saturation, and the amplitude path can be duty-cycled in burst operation. This advantage is meaningful in both cryogenic qubit controllers and MMW transmitters. However, it comes with a clear architectural constraint: the waveform is reconstructed from two paths, so amplitude–phase timing mismatch and bandwidth mismatch become first-order error mechanisms. When the paths are not aligned, the error does not appear as a pure $\Delta A$ or a pure $\Delta\phi$; it appears as a coupled envelope error that raises EVM and produces simultaneous rotation-angle and rotation-axis errors in qubit gates. Leakage at small amplitude codes and off-state spur suppression also matter more in polar than in Cartesian, because the amplitude path off condition is a primary operating mode during idle windows and back-off.

RF-DAC/DDS provides the most programmable waveform synthesis and the cleanest way to incorporate pulse shaping and reconstruction filtering, so it directly addresses the bandwidth and spectral-leakage bucket from Section II. In qubit control, this is valuable because shaped pulses reduce spectral spill that drives unwanted transitions, while per-cell trimming and calibration currents correct mismatch and suppress deterministic spur mechanisms. In MMW transmitters, RF-DAC architectures can support high-order modulation and, in some implementations, avoid lossy on-chip I/Q summation through spatial combining. The main limitation is signal accuracy from parasitic leakage and code-dependent distortion set a floor for low-amplitude constellation points, and ENOB/clock-jitter limits set the spur and noise floor. As a result, RF-DAC implementations depend strongly on segmentation, calibration, and spectral control to prevent leakage and code-dependent spurs from setting the EVM or gate-error floor.

Harmonic and subharmonic modulation relax LO distribution and frequency planning by deriving the RF carrier from a lower-frequency reference. This is attractive for scalable qubit controllers and wideband MMW/sub-THz systems because it reduces the number of high-frequency synthesizers and distribution paths that must be routed and isolated across multiple channels. The tradeoff is spectral cleanliness; harmonic extraction and subharmonic mixing introduce additional spur families and can tighten phase-noise requirements after frequency multiplication. Practical designs therefore rely on filtering and calibration of image and leakage terms. In MMW harmonic modulation, constant-impedance switching techniques such as common-node HOOK reduce switching-dependent loading and improve transient symmetry, which helps eye opening and array scalability, but spur control remains a defining constraint.

Despite recent advances, several challenges in both domains remain. In MMW systems, power efficiency at high peak to average power ratio (PAPR), calibration overhead, and LO distribution at sub-THz remain bottlenecks for scalable phased arrays. In cryogenic controllers, achieving <<1 mW per-qubit active power, while ensuring phase coherence across channels, and minimizing spur and LO leakage are critical for fault-tolerant quantum operation. Furthermore, both systems face stringent requirements for spectral containment, thermal management, and real-time calibration, especially under scaling demands. Therefore, future directions include: (1) hybrid polar–outphasing architectures for energy-aware waveform synthesis, (2) time-interleaved and multiplexed control for resource sharing, both for qubit array and MMW antenna arrays (3) cryo- and sub-THz-compatible packaging and 3D integration, and (4) embedding local calibration and algorithmic logic into ICs for autonomous operation. The co-evolution of MMW arrays and qubit-control hardware, guided by shared metrics and converging architectural techniques, is enabling more efficient and scalable waveform-synthesis platforms across both classical and quantum systems. These advances support a unified roadmap from quantum communication and sensing to high-capacity MMW and sub-THz connectivity.

## VI. Conclusion

This work has presented a unified comparative framework for evaluating direct digital-to-physical modulation techniques across MMW communication transmitters and cryogenic quantum controllers. Through architectural and circuit-level analysis of Cartesian I/Q, polar, RF-DAC, and harmonic/subharmonic modulators, we demonstrated how each topology trades off efficiency, calibration, spectral purity, signal integrity, and scalability. Notably, polar architecture offers superior energy efficiency with integrated amplitude/phase modulation, while harmonic-based designs provide broadband operation with reduced LO frequency constraints. RF-DACs enable high flexibility in frequency and envelope synthesis, and Cartesian systems maintain robust fidelity at the lowest power when operated with fixed LO and calibrated I/Q quadrature mixing. Importantly, this analysis reveals strong architectural and performance parallels between MMW and quantum control systems. Both domains require precise RF waveform generation, minimal spectral leakage, and efficient hardware scaling, whether for Gb/sec data transmission or high-fidelity quantum gate operations. Therefore, innovations in one field can directly inform and accelerate progress in the other.

### Acknowledgment

The authors thank Northeastern university for support. The authors also thank Dr. Jason Soric for the invitation. The chip fabrication for harmonic and subharmonic D-band has been supported under GlobalFoundries university shuttle program.

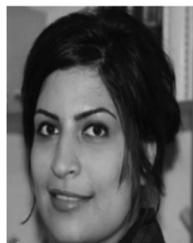

Najme Ebrahimi (Member, IEEE) received the Ph.D. degree from the University of California, San Diego, in 2017. From 2017 to 2020, she was a Postdoctoral Fellow at the University of Michigan, Ann Arbor. She is currently an Assistant Professor at Northeastern University, Boston, and was previously with the University of Florida. Her research interests include RF, mm-wave, and sub-THz ICs, wireless communications, IoT, and sensing. She is a recipient of the DARPA YFA (2021) and DARPA Director's Fellowship (2023), and serves on multiple IEEE TPCs, including RFIC, CICC, IMS, MILCOM, BCICTS, and the steering committee of IMS 2026 (RFTT).

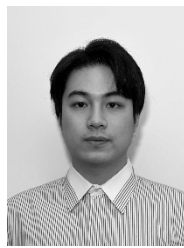

Haoling Li (Graduate Student Member, IEEE) received his B.E. degree in Electrical Engineering and Automation from Nanjing Tech University, Nanjing, China, in 2020, followed by an M.S. degree in Electrical and Computer

Engineering from Northeastern University, Boston, MA, USA, in 2023. Currently, he is pursuing a Ph.D. in Electrical Engineering at Northeastern University, Boston, MA, under the supervision of Dr. Ebrahimi in her research lab. His research focuses on RF, millimeter-wave, and terahertz circuits and systems.

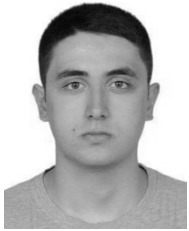

Gun Suer was born in Ankara, Turkey, on 29 November 2001. He received the B.S. degree in physics from Middle East Technical University (METU), Ankara, in 2023. He began pursuing the Ph.D. degree in physics at Northeastern University, Boston, MA, USA, in September 2023. His current research interests include superconducting qubits and quantum-limited amplifiers.

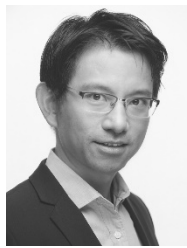

Kin Chung Fong (Member, IEEE) received the Ph.D. in Physics from The Ohio State University in 2008. From 2009 to 2013, he was a postdoctoral researcher at the Max Planck Institute for Quantum Optics and the California Institute of Technology. He is currently an Associate Professor at Northeastern University, Boston, and a Research Associate in the Department of Physics at Harvard University. From 2013 to 2024 he was Scientist I–Senior Scientist at Raytheon BBN Technologies. His research interests include quantum materials, single-photon detectors, quantum-noise-limited amplifiers, superconducting qubits, and electron hydrodynamics.

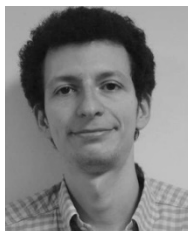

Leonardo Ranzani (Senior Member, IEEE) received the Ph.D. degree in Information Technology from Politecnico di Milano in 2008. From 2009 to 2015, he was a Research Associate at the University of Colorado Boulder and a Research Affiliate at the National Institute of Standards and Technology, Boulder. He is currently a Senior Scientist/Senior Principal Research Engineer at RTX BBN Technologies, Cambridge, MA. His research interests include superconducting and quantum microwave devices, including Josephson and kinetic-inductance parametric amplifiers, nonreciprocal microwave circuits, and qubit readout and control.